\documentclass[fleqn,11pt]{SelfArx} % Document font size and equations flushed left

\usepackage[english]{babel} % Specify a different language here - english by default

\usepackage{lipsum} % Required to insert dummy text. To be removed otherwise

\definecolor{color1}{RGB}{0,0,90} % Color of the article title and sections
\definecolor{color2}{RGB}{0,20,20} % Color of the boxes behind the abstract and headings

\usepackage{hyperref} % Required for hyperlinks

\hypersetup{
	hidelinks,
	colorlinks,
	breaklinks=true,
	urlcolor=color2,
	citecolor=color1,
	linkcolor=color1,
	bookmarksopen=false,
	pdftitle={Title},
	pdfauthor={Author},
}

\JournalInfo{Accepted to Astronomy Reports, 2025}
\Archive{}%DOI 10.1007/s11207-024-02365-0}

\PaperTitle{Do Planets Affect the Behavior of the Long-term Solar Activity?}
\PaperSubTitle{Do Planets Affect the Behavior of the Long-term Solar Activity?}

\Authors{M.M.~\ Katsova\textsuperscript{1}*, V.N.~\ Obridko\textsuperscript{2},  D.D.~\ Sokoloff\textsuperscript{2,3}, N.V.~\ Emelianov\textsuperscript{1}}

\affiliation{\textsuperscript{1}\textit{Sternberg Astronomical Institute of Lomonosov Moscow State University, Moscow, 119234, Russia}}
\affiliation{\textsuperscript{2}\textit{Pushkov Institute of Terrestrial Magnetism, Ionosphere and Radio
Wave Propagation, Kaluzhskoe shosse 4, Troitsk, 108840, Russia}}
\affiliation{\textsuperscript{3}\textit{Department of Physics, Lomonosov Moscow State University, Moscow, 119991, Russia}}
\affiliation{*\textbf{Corresponding author}: mkatsova@mail.ru}

\newcommand{\keywordname}{Keywords} % Defines the keywords heading name

\Abstract{
Solar activity is a process driven by many independent but interconnected phenomena. Although the 11-year cycle is the result of operation of the dynamo mechanism, the cause of longer secular variations is not clear. In search of such a cause, it was proposed to take into account the influence of the planetary system. In order to verify the idea, we consider the action of all planets in the solar system reduced to the effect of a single barycenter. The tidal force is decomposed into radial and meridional components. The radial tidal force is too small compared to the powerful radial gravity of the Sun. The meridional force is not compensated for by solar gravity and depends on latitude.
As the latitude of the barycenter changes quite slowly, the sign of this component changes over a characteristic time scale of about 5 years, during which the meridional acceleration constantly acts on the surface of the Sun. This could ultimately lead to speeds of several meters per second and, in principle, could significantly change the speeds of the meridional currents involved in generating the magnetic field. However, it turned out that the calculated speed variation does not agree with the observed periodicity of solar activity. Earlier, the relation was analyzed between the activity periods on solar-type stars and the rotation periods of exoplanets, and no correspondence was observed either. Thus, the planetary hypothesis as a cause of long-term modulation of solar activity is not confirmed.  
}

\begin{document}

\maketitle % Output the title and abstract box

%\tableofcontents % Output the contents section

\thispagestyle{empty} % Removes page numbering from the first page

%----------------------------------------------------------------------------------------
%	ARTICLE CONTENTS
%----------------------------------------------------------------------------------------

\section*{Introduction} % The \section*{} command stops section numbering

Solar activity is a complex process caused by many independent but interconnected phenomena. The cyclic activity of the Sun and stars is explained, first of all, by the generation of a magnetic field by a dynamo mechanism that transforms the energy of the poloidal field into the energy of the toroidal component as a result of differential rotation and mirror-asymmetric convection. This view has been supported for about a century and is presented in various books, reviews, articles, etc. (For example, see \cite{CS23} for a recent review of the history of the problem). The general idea was successfully applied to other late-type stars \cite{BB17}.

 We do not have to discuss the basic mechanisms of generation of magnetic fields on the Sun (dynamo processes). Note, however,  that the orbital period of Jupiter (11.86 years) is quite close to the duration of the solar cycle (about 11 years). The idea of the influence of Jupiter as a driving force of solar cyclic activity was put forward in the 19th century \cite{W59} and has survived in one form or another up to this day (e.g., \cite{J65, O20, Setal21}. The present-day observational astronomy rejects this idea in such a restricted form. The analysis of stellar activity in stellar systems with exoplanets shows that the planetary effects cannot be considered a general driver of the cyclic stellar activity \cite{Oetal22, Oetal24}. Indeed, there are systems with exoplanets, but no cyclic activity on the central star, and there is at least one late-type star with known cyclic activity where no exoplanets were found. 

\cite{Oetal22, Oetal24} analyzed new data on the previously considered solar-type stars with identified cycles and added data on long-term variability on two more solar-type G stars and 15 cooler M dwarfs with planets. If the cyclic activity of a star is determined by a strong tidal influence of the planet, then the stellar activity cycle should be synchronized with the planet's orbital period. The gravitational effect of planets on their parent stars was calculated. The results obtained confirm the earlier conclusion that exoplanets do not influence the formation of the stellar cycle. 

However, we cannot say that the planets do not affect in any way the cyclic activity on stars. Moreover, it seems quite plausible that the magnetic activity of a member of a close binary system may be modulated by the influence of its companion, e.g., by the reflection effect, e.g. \cite{Metal02}. Identifying some of the planetary effects of solar activity could be an important step in solar activity research. 

Some authors, e.g. \cite{Setal21}, link the planetary effects to long-term variations of the geomagnetic index, {\bf aa}, which is not a direct metric of solar activity, such as the Gleissberg cycle or a longer cycle. The argument presented by \cite{Oetal22} does not exclude this option. \cite{O20} suggests that the Global Minima of solar activity correspond to the periods when the barycenter of the solar planetary system is located at a distance larger than two solar radii from the center of the Sun. \cite{Oetal24} examined the change in the position of the barycenter of the solar system relative to the center of the Sun over 420 years and found that the data of celestial mechanics do not confirm this suggestion.  In this paper, they show that the structure of distances from the barycenter to the center of the Sun during the Maunder minimum from 1645 to 1710 does not differ from what is observed in our epoch from 1976 to 2019. An important difference between the solar cycle and the barycenter motion can be presented as follows. The latter is much more stable than the solar cycle. A comparison of these data with the 120-year SSN (sunspot number) series as the index of solar activity shows that they are not synchronized \cite{Oetal24}. 

\cite{Weissetal2023} also showed the lack of synchronization in the solar system. This result strongly supports randomly perturbed dynamo models with small intercycle memory.

The key point of our analysis is that the coincidence in time-scales between the magnetic activity and the planetary motions is insufficient to speculate about a physical connection between them. One has to demonstrate some regular link between the phases of the phenomena (e.g., a regular lag between the force and its effect). This is a key novelty of the present paper compared to the previous one \cite{Oetal24}.

\begin{figure}[h!]
        \centering 
        \includegraphics[width=0.45\textwidth]{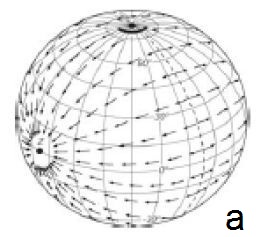}
        \includegraphics[width=0.45\textwidth]{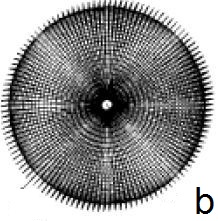}
        \caption{Distribution of horizontal tidal forces from the single planet located above the equator of the star(a). The structure of the vertical tidal forces on the Sun in September 2005 (b).}
       \label{F1}
       \end{figure}

\section{Calculating tidal forces}

Let us consider the action of all planets in the solar system reduced to the effect of a single barycenter.

For a single planet orbiting the Sun in the equatorial plane, the tidal potential $V$ is

\begin{equation}
    V=- \frac{\gamma M r^2}{2R^3}(3 \cos^2 \phi -1),
    \label{Eq1}
\end{equation}
where $\gamma$ is the gravitational constant, $M$ is the mass of the planet, $R$ is the distance between the Sun's centers and the planet, $r$ is the distance from the solar center (for the cases of interest, here $r \approx R_\odot \approx 7 \times 10^8$ m), and $\phi$ is the latitude on the solar surface. The gravitational potentials for several planets are additive.

The tidal force is decomposed into radial and meridional components. The radial component is unimportant to us, being negligible compared to the radial gravitational attraction of the Sun. We do not consider here the third component of the tidal force orthogonal to the radial and meridional coordinates, because the solar activity data are averaged in this direction. More precisely, we calculate the force per unit mass, i.e. the acceleration.

The meridional force

\begin{equation}
    F_M=-\frac{1}{r} \frac{\partial V}{\partial \phi} = \frac{2G}{r} \sin 2 \phi,
    \label{Eq2}
\end{equation}
where $G= \frac{3}{4} \gamma M r^2/R^3$, is not compensated for by the solar attraction and depends on the barycenter latitude. The meridional acceleration is the force per unit of mass. In order to avoid complicated notation, we also use $F_M$ for acceleration.  The latter varies quite slowly within the time scale of about 5 yrs (half of the Jupiter orbital period), which yields a meridional acceleration of the same time scale acting on the solar surface. This acceleration can produce velocities of about several microseconds per second, which may, in principle, cause substantial modifications in the solar meridional circulation \cite{Getal12, G13, Netal21} that play a significant role in solar cyclic activity explaining the effects of the North-South asymmetry \cite{BB13}. 

A particular modification of the meridional circulation depends on solar magnetohydrodynamic (MHD) flows and can differ in different solar dynamo models. Including the force in all known solar dynamo models to isolate the effect seems impractical. However, if the effect is significant, we can expect a regular relationship between the force variation and the time behavior of the solar activity. The search for such a pronounced relationship is our goal.

\begin{figure}[h!]
        \centering 
        \includegraphics[width=0.85\textwidth]{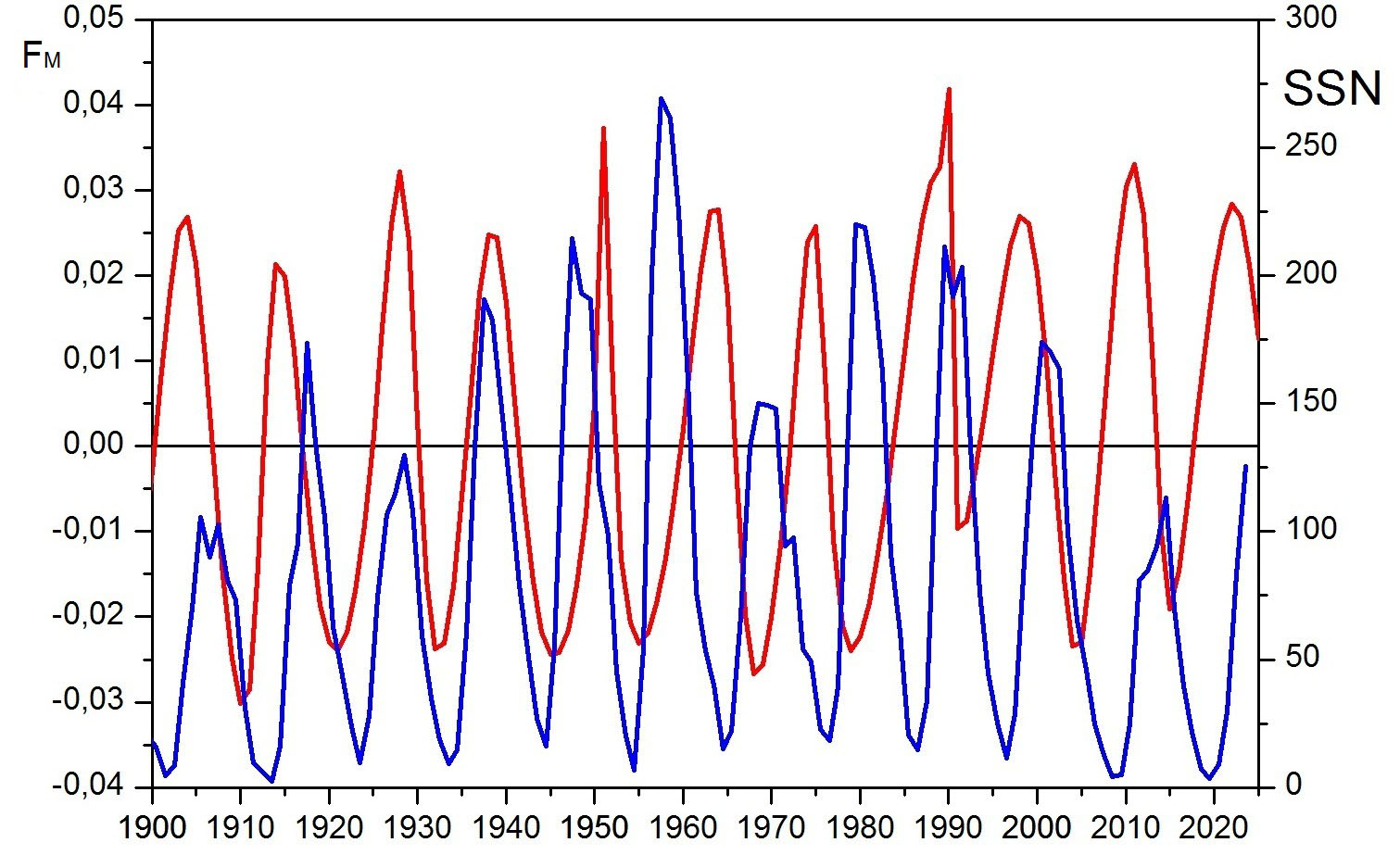}        
		\caption{Meridional tidal acceleration $F_M$ (force per unit mass) measured in m/s$^{2}$ on the surface of the Sun (red) and sunspot numbers (SSN, blue) over the past 120 years.
 }
        \label{F2}
    \end{figure}

\section{Results: no clear link between the planetary motion and solar activity}

The use of these formulas is justified when a single planet is analyzed or if we want to analyze a short-term database of solar activity. This approach was successfully applied by \cite{Getal09} and later, according to the same principle, by \cite{Oetal22}, \cite{Oetal24}. However, when using the contribution of all planets, the velocity field on the Sun becomes too complex even over a short time interval, and it is difficult to use it for analyzing the periodicity.
   
Fig 1a shows the distribution of horizontal tidal forces from a single planet located above the equator of the star. In the case of the Sun, with several planets, the tidal forces depend on the distance and position of the planets, and these parameters depend on time. In this case, the structure of the forces becomes much more complex. Fig. 1b shows the structure of the vertical tidal forces on the Sun in September 2005. (Part of the pictures are taken from the paper by \cite{Getal09}.

Direct use of formulas (1) and (2) for each of the planets of the solar system in September 2005 and analysis of their total effect yields the picture shown in Figure 1b. Such a picture changes with time and is difficult to use in a  long-term statistical analysis. Therefore, we refused from using formulas (1) and (2) separately for each planet and took into account the total tidal effect using the barycenter of the solar system.

 For this purpose, an algorithm for calculating the ephemerides of all planets in the solar system is used. In our earlier work \cite{Getal09} we used the JPL Planetary and Lunar Ephemrides version
DE406, which specifies the planet positions at any time covered by their time span in the ICRS Earth-centered reference frame. 
All nine planets were used including Pluto (as specified in the Ephemerides, excluding only the Asteroids). Instead of the Earth alone, the Earth-Moon barycenter is used, with the combined mass of the Earth and the Moon \cite{Stand08}. 

Later, this version was improved by adding asteroids (versions DE430 and DE431). DE430 and DE431 differ in their integrated time span and lunar dynamical modeling.    
The dynamical model for DE430 includes a damping term between the Moon's liquid core and solid mantle, which best fits the range data but is not suitable for forward integration over timescales longer than a few centuries. For ephemeris DE431 is similar to DE430 but was fit without the core/mantle damping term, so the lunar orbit is less accurate than in DE430 for times near the current epoch, but is more suitable for times more than a few centuries in the past. DE431 is a longer integration (covering years -13,200 to + 17,191) than DE430 (covering years 1150 to 2650). The positions of the barycenter of the solar system relative to the center of the Sun were calculated using the DE431 ephemerides described by \cite{Fetal14}.

To perform this verification, we used the coordinates of the north pole of the Sun from \cite{Aetal18}. For other approaches to including the tidal force in the context of solar activity investigation, see \cite{Cetal23}, who are using the spectral approach proposed by \cite{K04, K09} .

To start with, we compared data for the past 120 years (Fig.~\ref{F2}). Of course, both parameters demonstrate a clearly pronounced periodicity. Some maxima of the blue curve are close to the maxima of the red curve (for example, 1930-1940, cycles 16-17), whereas the other are close to its minima (cycles 23-24). This is not surprising since the rotation period of the barycenter during this time interval is 11.8 years, which is very close to the rotation period of Jupiter, and the mean duration of the solar cycle over the past 120 years is close to 10.6 years. It is clear that if at some point the extremes of the curves coincide, after 5-7 cycles the extremes will be in antiphase. Thus, we conclude that the phase relationship between celestial data and solar activity on the scale of the Schwabe cycle (11 years) seems questionable.

\begin{figure}[t]
        \centering 
        \includegraphics[width=0.85\textwidth]{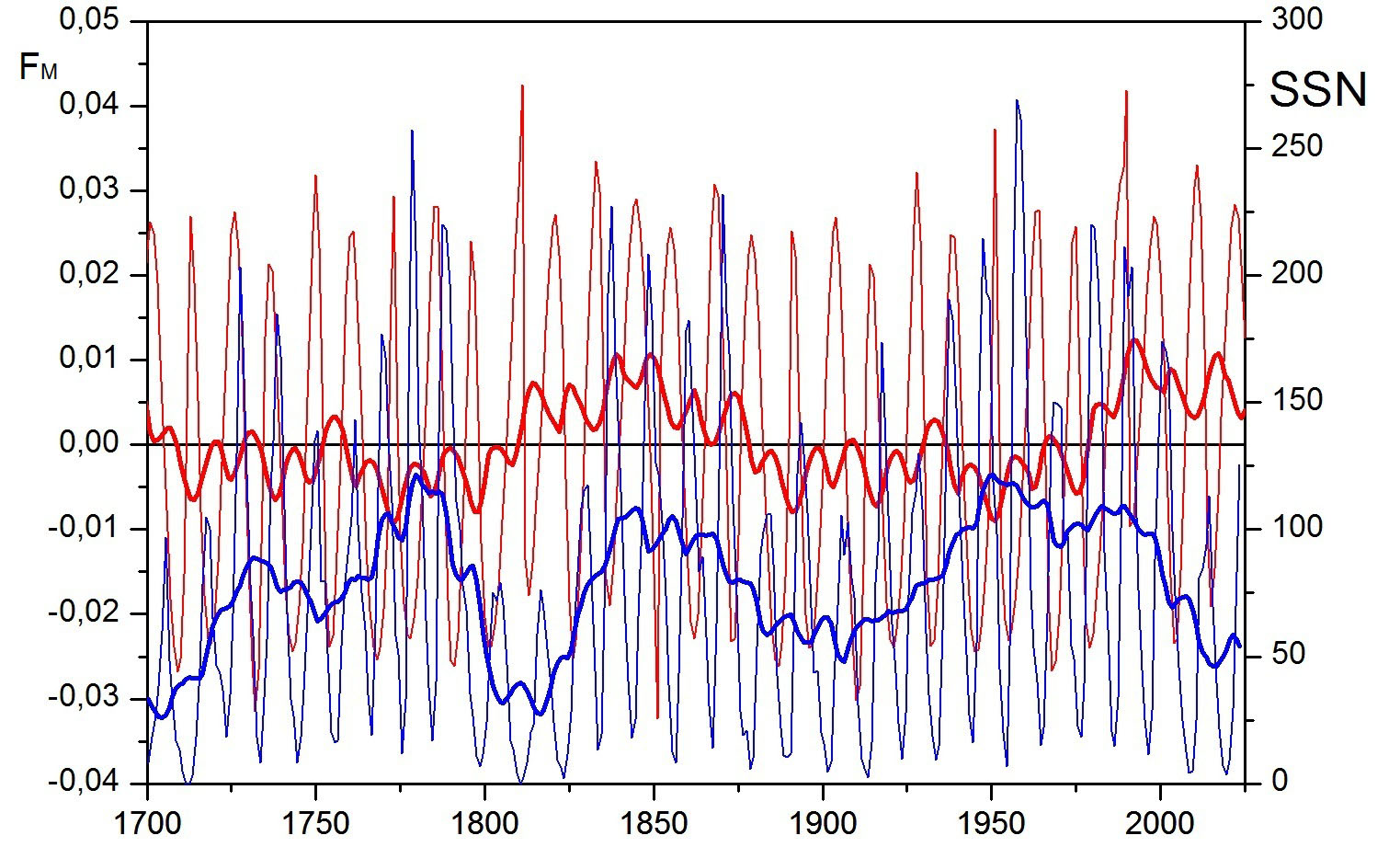}
        \caption{Meridional tidal acceleration $F_M$ measured in m/s$^{2}$ on the surface of the Sun (thin red line) and sunspot numbers (SSN, thin blue line) over the past 320 years. The thick lines show the corresponding 20-yrs averaging.
 }
        \label{F3}
    \end{figure}

As the interval of 120 years shown in Fig.~\ref{F2} may seem too short, we also compared the data of celestial mechanics and solar activity for the past 320 yrs, i.e. over the interval where the instrumental sunspot data are more or less reliable (Fig.~\ref{F3}). Again, we see no clear phase relation of the signals on the time scales of the Schwabe cycle. However, the plot seems more suitable for searching for phenomena with a timescale of about 100 years, such as the Gleissberg cycle.

\begin{figure}[h!]
        \centering 
       \includegraphics[width=0.85\textwidth]{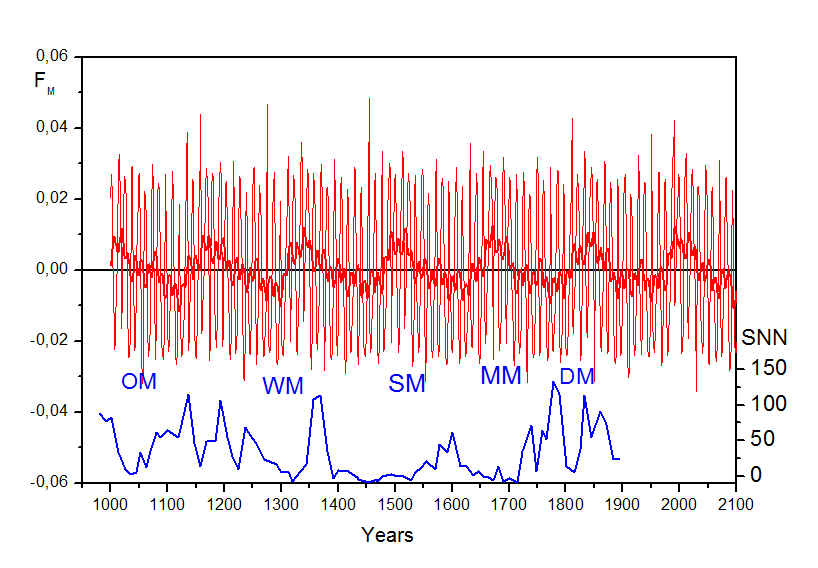}
        \caption{Meridional acceleration (thin red line) and its 20-year average (thick red line) compared to reconstructed SSN data (blue). The legends OM, WM, SM, MM DM mark the epochs of special behavior of solar activity: the Oort Minimum, Wolf Minimum, Sp\"orer Minimum, Maunder Minimum, and Dalton Minimum.}
\label{F4}
    \end{figure}

To compare the planetary hypothesis with longer-term features of solar activity evolution, we compare the evolution of the meridional force with the reconstructed sunspot data for the previous millennium (Fig.~\ref{F4}).
 
As the time resolution of the reconstructed millennial data is obviously lower than that of the telescopic observations used in the previous figures (e.g., because the atmospheric lifetime of the isotopes in question is comparable with the Schwabe cycle), we do not discuss here possible correlations between the solar activity data and tidal forces. As for possible correlations with the solar Grand Minima, we see no clear relationship between both records.

To try an alternative approach, we analyzed the speed (m/s) that meridional flows would reach in a year given the previously obtained acceleration ($F_M$). Then, we estimated the changes in this speed over 1000 years and compared them with the solar activity index (SSN) reconstructed from annual $^{14}C$ data over the past millennium \cite{Uetal21, Uetal2025} to find out that there is no clear relationship (Fig.~\ref{F4}). 

When comparing the tidal force and solar activity, we have to take into account that the quality of modern solar-activity data is obviously better than the quality of historical data. 

\begin{figure}[h!]
        \centering 
\includegraphics[width=0.85\textwidth]{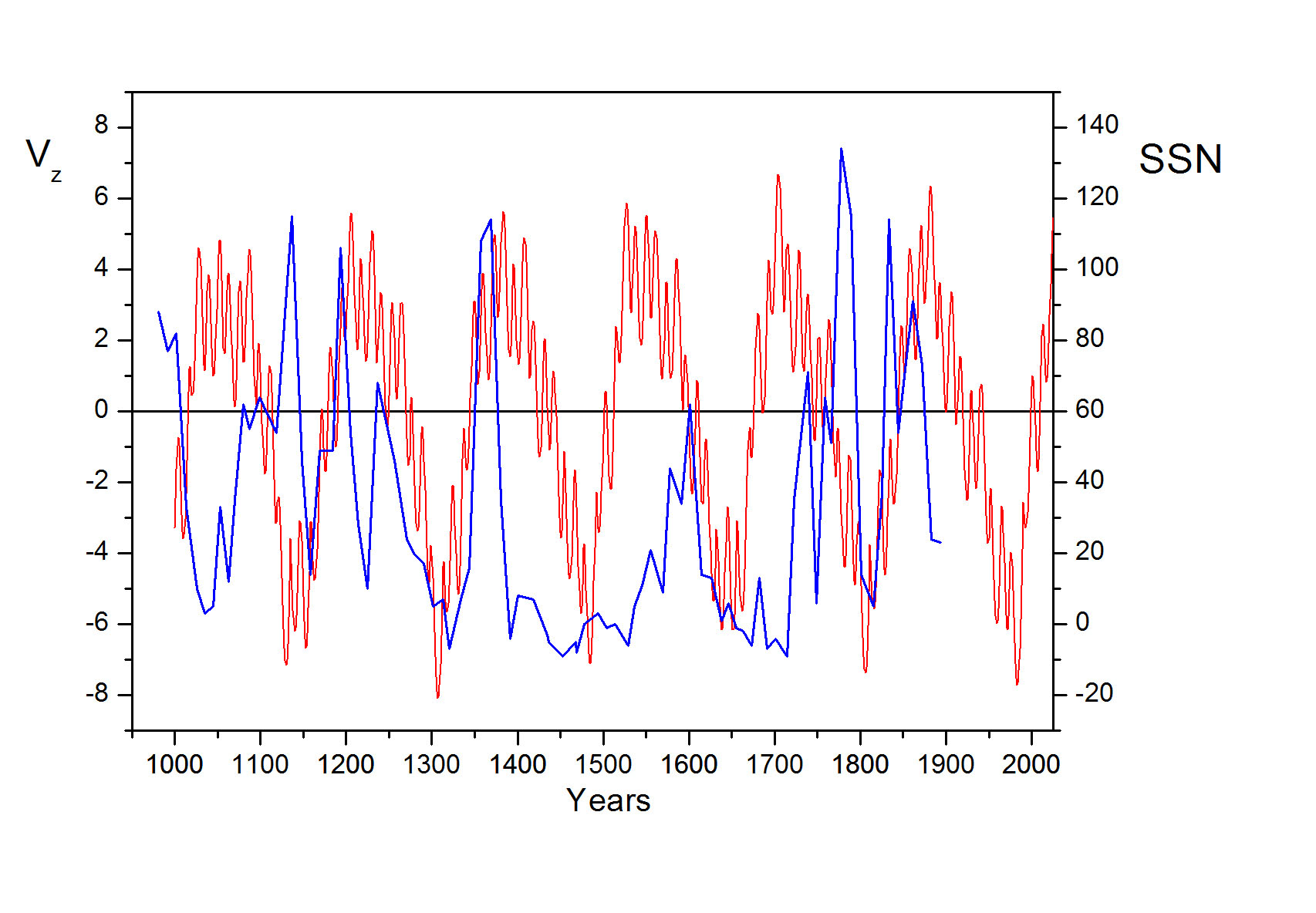}
        \caption{Accumulated meridional velocity $V_M$ (red) versus the reconstructed SSN data (blue) over the past millennium.
 }
    
\label{F5}
    \end{figure}

Perhaps, the most impressive feature in Figs.~\ref{F4} and \ref{F5} is a pronounced oscillation with a timescale of about 178 years in the celestial mechanics data, reported by \cite{J65}. The periodic nature of this oscillation seems to be more pronounced than in Figs.~\ref{F2} - \ref{F3}. However, we see no clear relationship between this periodicity and variations in the solar activity data. We can theoretically admit that the quality of the reconstructed SSN data is insufficient to isolate such a correlation and connect it with the Gleissberg cycle or other cyclic features in the solar activity data reported in scientific literature; however such a conclusion would require a basic improvement in the reconstructed SSN data. 
 
\section{Conclusion and Discussion}

We must emphasize once again that we do not reject the possibility of isolating the influence of planetary motion and tidal effects on the formation and development of solar cyclic activity. The tidal forces are strong enough to significantly affect the meridional circulation in the solar convection zone. However, we are more interested in the really observed effects than in the possible ones. We should note that a pure coincidence of time scales in the sunspot data and celestial mechanics is insufficient to talk about a physical connection between the processes. A physical mechanism or at least a stable phase relationship between the oscillations is needed. Obviously, the promoters of the idea of planetary effects in solar activity have to suggest a physical mechanism. We are quite reserved with respect to this point.  

We tried our best to find phase relationships between the planetary effects and solar activity. Of course, in addition to the regular Schwabe cycle, solar activity records show pronounced random features. Thus, there is a possibility that these random processes may, at one time or another, imitate some features of deterministic planetary motion.The point is that we do not see this in the available observations. One might think that an extension of the observational database and/or new physical ideas about the influence of planetary motion on the solar MHD could give rise to a new approach to the topic; however, we do not see such possibilities in the current state of knowledge.

There is also another point worth mentioning. \cite{Oetal22, Oetal24} recently presented a list of exoplanetary systems, where the tidal forcing of the central star is much stronger than that observed in the solar system, whereas the known magnetic activity of the central star does not show a relationship between the main period of activity and the orbital period similar to that in the solar system. This fact supports our conclusion that the relationship between planetary effects and cyclic stellar activity is weaker than might be expected.

In general, the situation is really complicated. Planets can either disrupt the periodicity or stabilize it. But they cannot create it, and we do not see synchronization anywhere. In addition, there is no reason to believe that the dynamo cannot create periodicity of any period without outside influence. The unique feature of the Solar System is the proximity of Jupiter's rotation period to the 11-year cycle. But here,  there are also certain difficulties; there is no complete synchronization, and the mechanism of interaction of internal processes with an external force is not clear.

Of course, the closeness of the rotation periods of the barycenter and the 11-year cycle is suggestive, especially as the length of a solar cycle (usually called 11-year) is very uncertain.  E.g., from Cycle 14 to Cycle 24, the mean cycle length was 10.719$\pm$0.721 years, and from Cycle 8 to Cycle 24, it was 10.946$\pm$0.83 years. These values are obtained using smoothed sunspot numbers from 
\url{https://sidc.be/SILSO/cyclesminmax},  version 2. Other data or other indices may give slightly different values, but in all cases, the spread is quite large. By selecting necessary values and including the chosen planets in calculation (e.g., only Jupiter and Saturn), we can confirm the period of 10-11 years and, according to the laws of amplitude modulation, obtain additional periods from 80 to 170 years. However, this is impossible to verify experimentally. We have reliable unrecovered data only since the mid-19th century.

Another point is that, at best, the planetary hypothesis deals only with periods. It does not attempt to explain all the other important properties of solar activity (and above all the fact that the physical cycle is not an 11-year cycle, but a magnetic 22-year cycle with a regular change in the characteristics of the magnetic field). At the same time, all these properties receive their logical explanation in the dynamo mechanism.

Thus, our general conclusion is the absence of synchronization between the quasi-periodic main 11-year cycle of solar activity and the strictly periodic rotation of the barycenter of the planets. This conclusion agrees with the results of two earlier works \cite{Oetal22, Oetal24} that proved the absence of synchronization in systems with exoplanets. Such synchronization would inevitably occur if the planetary influence were the main mechanism of solar and stellar activity. Thus, there is no doubt that solar activity is an endogenous process determined by the solar dynamo.

However, the similarity between the main rotation period of the barycenter and the length of the solar cycle may lead to variations in the height and duration of individual cycles because of weak beats. The presence of a weak connection does not ensure full synchronization and preservation of phase shift.

Tangential forces can, in principle, slightly change the meridional currents and, accordingly, change the height of the following solar cycle. At the same time, the effect of tidal forces on the solar and stellar dynamo is more complex than the modulation of the meridional circulation discussed by the authors. This can, in principle,  explain deviations from direct synchronization with the influence of tidal forces. For example, \cite{Kletal2023} showed that the tidal interaction leads to modulation of turbulent convection and inertial waves and hence to variations in the alpha effect. Such an interaction may be rather complex. The influence of tidal interaction on the dynamo 
(so far, for stronger interactions in stellar systems) is studied using MHD simulations in \cite{Astbar2025}. The long-term periodicity is a separate issue. Of course, the 178-year cycle is indeed close to the so-called Suess cycle. However, this cycle itself was discovered using a too short data set, and its reliability is low. In general, we cannot rule out that the planetary influence may weakly modulate the height of the solar cycle. On the other hand, these effects may be generated directly by the solar dynamo mechanism with weakly varying parameters included in the calculations without any external causes.

\subsection*{Conflict of interest} The authors declare that they have no conflict of interest.

\subsection*{Funding}
The study was carried out under the state assignment of Lomonosov Moscow State University.

\subsection*{Data Availability}
The sunspot data were taken from WDC-SILSO, Royal Observatory of Belgium, Brussels \url{https://sidc.be/SILSO/datafiles}.

\subsection*{Author Contributions} All authors contributed to the conception and design of the study.

%Material preparation and data collection were performed by M.Katsova.

%The calculations of the barycenter and accelerations were performed by N.Emelianovv.

%The juxtaposition with the solar data was performed by V.Obridko.

%D.Sokoloff revised previous versions of the manuscript. All authors read and approved the final manuscript.

\bibliographystyle{unsrt}

\bibliography{cycle}

\begin{thebibliography}{10}

\bibitem{CS23}
Paul {Charbonneau} and Dmitry {Sokoloff}.
\newblock {Evolution of Solar and Stellar Dynamo Theory}.
\newblock {\em \ssr}, 219(5):35, August 2023.

\bibitem{BB17}
Allan~Sacha {Brun} and Matthew~K. {Browning}.
\newblock {Magnetism, dynamo action and the solar-stellar connection}.
\newblock {\em Living Reviews in Solar Physics}, 14(1):4, December 2017.

\bibitem{W59}
R.~{Wolf}.
\newblock {Extract of a Letter to Mr. Carrington}.
\newblock {\em \mnras}, 19:85--86, January 1859.

\bibitem{J65}
Paul~D. {Jose}.
\newblock {Sun's motion and sunspots}.
\newblock {\em Astron. J.}, 70:193, April 1965.

\bibitem{O20}
V.~P. {Okhlopkov}.
\newblock {11-Year Index of Linear Configurations of Venus, Earth, and Jupiter
  and Solar Activity}.
\newblock {\em Geomagnetism and Aeronomy}, 60(3):381--390, June 2020.

\bibitem{Setal21}
F.~{Stefani}, R.~{Stepanov}, and T.~{Weier}.
\newblock {Shaken and Stirred: When Bond Meets Suess-de Vries and
  Gnevyshev-Ohl}.
\newblock {\em \solphys}, 296(6):88, June 2021.

\bibitem{Oetal22}
V.~N. {Obridko}, M.~M. {Katsova}, and D.~D. {Sokoloff}.
\newblock {Solar and stellar activity cycles - no synchronization with
  exoplanets}.
\newblock {\em \mnras}, 516(1):1251--1255, October 2022.

\bibitem{Oetal24}
V.~N. {Obridko}, M.~M. {Katsova}, D.~D. {Sokoloff}, and N.~V. {Emelianov}.
\newblock {Is There a Synchronizing Influence of Planets on Solar and Stellar
  Cyclic Activity?}
\newblock {\em \solphys}, 299(9):124, September 2024.

\bibitem{Metal02}
D.~{Moss}, N.~{Piskunov}, and D.~{Sokoloff}.
\newblock {Nonaxisymmetric cool spot distributions and dynamo action in close
  binaries}.
\newblock {\em \aap}, 396:885--893, December 2002.

\bibitem{Weissetal2023}
E.~{Weisshaar}, R.H. {Cameron}, and M.~{Schüssler}.
\newblock No evidence for synchronization of the solar cycle by a “clock”.
\newblock {\em \aap}, 671:A87, 2023.

\bibitem{Getal12}
V.~N. {Obridko}, Yu.~A. {Nagovitsyn}, and Katya {Georgieva}.
\newblock {The Unusual Sunspot Minimum: Challenge to the Solar Dynamo Theory}.
\newblock In Vladimir~N. {Obridko}, Katya {Georgieva}, and Yury~A.
  {Nagovitsyn}, editors, {\em The Sun: New Challenges}, volume~30 of {\em
  Astrophysics and Space Science Proceedings}, page~1, January 2012.

\bibitem{G13}
Katya {Georgieva}.
\newblock {Space Weather and Space Climate{\textemdash}What the Look from the
  Earth Tells Us About the Sun}.
\newblock In Jean-Pierre {Rozelot} and Coralie (Eds.~) {Neiner}, editors, {\em
  Lecture Notes in Physics}, volume 857, page~53. Berlin Springer Verlag, 2013.

\bibitem{Netal21}
Dibyendu {Nandy}, Petrus C.~H. {Martens}, Vladimir {Obridko}, Soumyaranjan
  {Dash}, and Katya {Georgieva}.
\newblock {Solar evolution and extrema: current state of understanding of
  long-term solar variability and its planetary impacts}.
\newblock {\em Progress in Earth and Planetary Science}, 8(1):40, December
  2021.

\bibitem{BB13}
Bernadett {Belucz} and Mausumi {Dikpati}.
\newblock {Role of Asymmetric Meridional Circulation in Producing North-South
  Asymmetry in a Solar Cycle Dynamo Model}.
\newblock {\em Astrophys. J.}, 779(1):4, December 2013.

\bibitem{Getal09}
K.~{Georgieva}, P.A. {Semi}, B.~{Kirov}, V.~{Obridko}, and B.D. {Shelting}.
\newblock {Planetary tidal effects on solar activity}.
\newblock In {\em Proc. All Russian Workshop on the Solar Physics "An Year of
  the Astronomy: Star and Solar-terrestrial Physics"}, pages 117--120, June
  2009.

\bibitem{Stand08}
E.M. {Standish}.
\newblock {Planetary and Lunar Ephemerides: testing alternate gravitational
  theories}.
\newblock In {\em Recent Developments in Gravitation and Cosmology: 3rd Mexican
  Meeting on Mathematicaland Experimental Physics. AIP Conference Proceedings},
  volume 977, pages 254--263, 2008.

\bibitem{Fetal14}
W.~M. {Folkner}, J.~G. {Williams}, D.~H. {Boggs}, R.~S. {Park}, and
  P.~{Kuchynka}.
\newblock {The Planetary and Lunar Ephemerides DE430 and DE431}.
\newblock {\em Interplanetary Network Progress Report}, 42-196:1--81, February
  2014.

\bibitem{Aetal18}
B.~A. {Archinal}, C.~H. {Acton}, M.~F. {A'Hearn}, A.~{Conrad}, G.~J.
  {Consolmagno}, T.~{Duxbury}, D.~{Hestroffer}, J.~L. {Hilton}, R.~L. {Kirk},
  S.~A. {Klioner}, D.~{McCarthy}, K.~{Meech}, J.~{Oberst}, J.~{Ping}, P.~K.
  {Seidelmann}, D.~J. {Tholen}, P.~C. {Thomas}, and I.~P. {Williams}.
\newblock {Report of the IAU Working Group on Cartographic Coordinates and
  Rotational Elements: 2015}.
\newblock {\em Celestial Mechanics and Dynamical Astronomy}, 130(3):22, March
  2018.

\bibitem{Cetal23}
Rodolfo~G. {Cionco}, Sergey~M. {Kudryavtsev}, and Willie W.~H. {Soon}.
\newblock {Tidal Forcing on the Sun and the 11-Year Solar-Activity Cycle}.
\newblock {\em \solphys}, 298(5):70, May 2023.

\bibitem{K04}
S.~M. {Kudryavtsev}.
\newblock {Improved harmonic development of the Earth tide-generating
  potential}.
\newblock {\em Journal of Geodesy}, 77(12):829--838, June 2004.

\bibitem{K09}
S.~M. {Kudryavtsev}.
\newblock {Long-term harmonic development of lunar ephemeris}.
\newblock {\em \aap}, 471(3):1069--1075, September 2007.

\bibitem{Uetal21}
I.~G. {Usoskin}, S.~K. {Solanki}, N.~A. {Krivova}, B.~{Hofer}, G.~A.
  {Kovaltsov}, L.~{Wacker}, N.~{Brehm}, and B.~{Kromer}.
\newblock {Solar cyclic activity over the last millennium reconstructed from
  annual $^{14}$C data}.
\newblock {\em \aap}, 649:A141, May 2021.

\bibitem{Uetal2025}
I.~{Usoskin}, T~{Chatzistergos}, S.K. {Solanki}, N~{Krivova}, G.~{Kovaltsov},
  N.~{Brehm}, M.~{Christl}, and L.~{Wacker}.
\newblock Sunspot cycles for the first millennium bc reconstructed from
  radiocarbon.
\newblock {\em \aap}, 698:A182, 2025.

\bibitem{Kletal2023}
M.~Klevs, F.~Stefani, and L.~Jouve.
\newblock A synchronized two-dimensional $\alpha$-$\omega$ model of the solar
  dynamo.
\newblock {\em Solar Physics}, 298, July 2023.

\bibitem{Astbar2025}
A.~Astoul and A.J. Barker.
\newblock Interplay between tidal flows and magnetic fields in non-linear
  simulations of stellar and planetary convective envelopes.
\newblock {\em MNRAS}, 541:1575–1599, June 2025.

\end{thebibliography}

\end{document}